\documentclass[11pt]{cernrep}
\usepackage{graphicx}
\usepackage{cite,./mcite}

\usepackage[latin1]{inputenc}
\usepackage{rotating}
\usepackage{exscale}
\usepackage{latexsym}
\DeclareGraphicsExtensions{.eps.gz,.eps,.ps.gz,.ps,.pdf}

\def\Journal#1#2#3#4{{#1}{\bf \ #2}, #3 (#4)}
\def\CPC{Comput. Phys. Commun.}

\def\EPJ{Eur. Phys. J. {\bf C}}

\def\NPB{Nucl. Phys. {\bf B}}

\def\PLB{Phys. Lett. {\bf B}}
\def\PRL{Phys. Rev. Lett.}
\def\PRD{Phys. Rev. {\bf D}}
\def\ZPC{Z. Phys. {\bf C}}

\def\RMP{Rev. Mod. Phys.}
\def\SNP{Sov. J. Nucl. Phys. }
\def\JETP{Sov. Phys. JETP }
\def\PRP{Phys. Rept. }
\def\NPA{Nucl. Phys. {\bf A}}
\def\APPB{Acta Phys.\ Polon. {\bf B}}
 
\def\JPG{J. Phys. {\bf G}}

\newcommand{\lwig}{\mbox{\,\raisebox{.3ex}
    {$<$}$\!\!\!\!\!$\raisebox{-.9ex}{$\sim$}\,}}
\newcommand{\gwig}{\mbox{\,\raisebox{.3ex}
    {$>$}$\!\!\!\!\!$\raisebox{-.9ex}{$\sim$}}\,}

\newcommand{\iai}{I\overline{I}}

\newcommand{\ii}{{\rm i}}

\newcommand{\xpr}{{x^\prime}}

\newcommand{\xbj}{x_{\rm Bj}}

\newcommand{\sidp}{\sigma_{\rm \tiny DP}}
\newcommand{\rav}{\langle\rho\rangle}
\newcommand{\vl}{\mathbf{l}}
\newcommand{\vr}{\mathbf{r}}

\begin{document}
\title{{\normalsize\rightline{DESY
05-125}\rightline{\lowercase{hep-ph/0507160} }}
\vspace{-5ex}
Instanton-Induced Processes \\ An Overview\footnote{\hspace{2ex}Invited
plenary talk presented at the CERN - DESY Workshop 2004/2005, {\it HERA and
the LHC}.}}
\author{F. Schrempp}
\institute{Deutsches Elektronen-Synchrotron DESY, Hamburg, Germany }
\maketitle
\begin{abstract}
A first part of this review is devoted to a summary of our extensive studies
of the discovery potential for instanton\,($I$)-induced, deep-inelastic
processes at HERA. Included are some key issues about $I$-perturbation theory,
an exploitation of crucial lattice constraints and a status report about the
recent $I$-search results by the HERA collaborations H1 and ZEUS in relation to
our predictions. Next follows a brief outline of an ongoing project concerning a
broad exploration of the discovery potential for hard instanton processes at the
LHC. I then turn to an overview of our work on high-energy processes, involving
larger-sized instantons. I shall mainly focus on the phenomenon of saturation at
small Bjorken-$x$ from an instanton perspective. In such a framework, the
saturation scale is associated with the conspicuous average instanton size,
$\rav \sim 0.5$ fm, as known from lattice simulations. A further main result
is the intriguing identification of the ``Colour Glass Condensate" with the
QCD  {\it sphaleron} state.
\end{abstract}
\section{Setting the stage} 
{\it Instantons} represent a basic non-perturbative aspect of
non-abelian gauge theories like QCD. They were theoretically discovered and
first studied by Belavin {\it et al.}~\cite{bpst} and `t
Hooft~\cite{th}, about 30 years ago. 

Due to their rich vacuum structure, QCD and similar theories
include topologically non-trivial fluctuations of the gauge fields,
which in general carry a conserved, integer topological charge $Q$.
Instantons ($Q=+1$) and anti-instantons  ($Q=-1$) represent the
simplest building blocks of topologically non-trivial vacuum
structure. They are explicit solutions of the euclidean
\begin{figure}
\begin{center}
\includegraphics*[width=5.8cm]{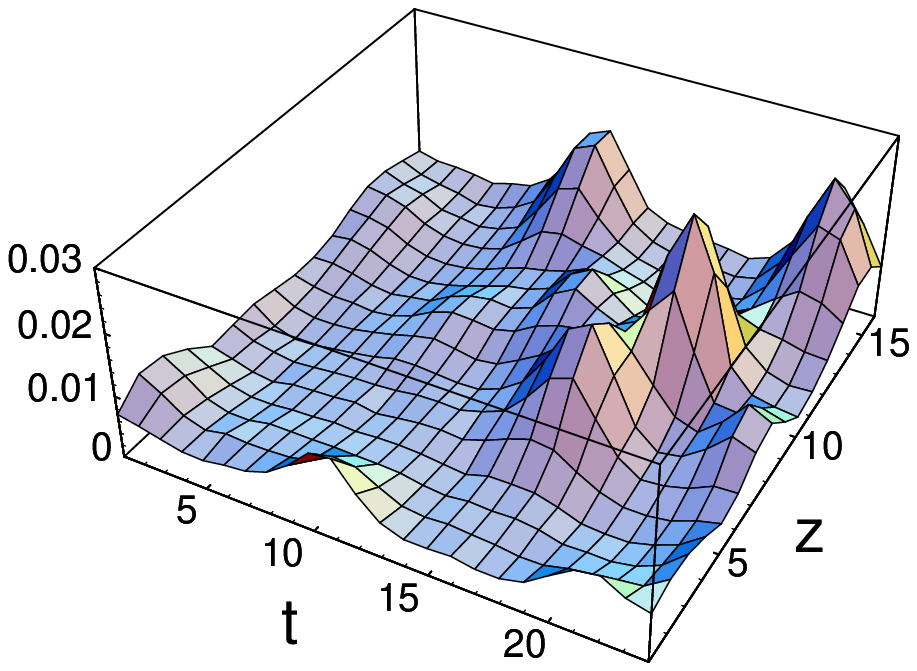}\hspace{1.5cm}
\includegraphics*[width=5.8cm]{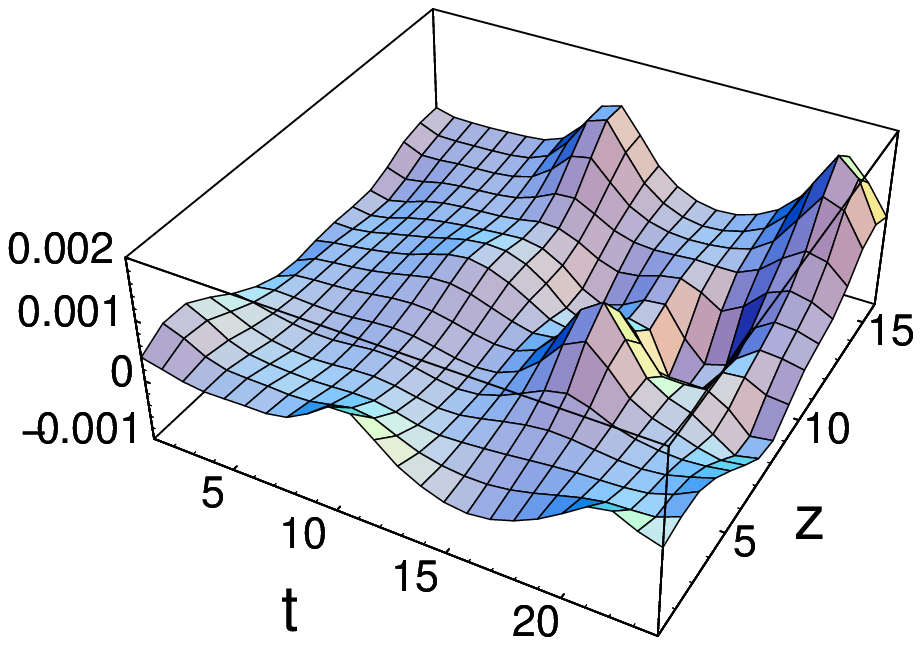}
\caption[dum]{Contribution from three instantons ($Q=+1$) and two
anti-instantons ($Q=-1$) to the Lagrangian (left)) and the topological
charge
density (right) in a lattice simulation~\cite{chu} (after cooling). The
euclidean coordinates x and y are kept fixed while the dependence on z and
t is displayed.}
\end{center}
\end{figure}
field equations in four dimensions~\cite{bpst}. They are known to play an
important r{\^o}le in the transition region between a partonic and a
hadronic description of strong interactions~\cite{ssh}. Yet, despite
substantial theoretical evidence  for the importance of instantons in
chiral symmetry breaking and hadron spectroscopy,  their direct
experimental verification is lacking until now.

It turns out, however, that a characteristic {\it short distance}
manifestation of instantons can be exploited~\cite{rs1} for an
experimental search: Instantons induce certain (hard) processes that
are forbidden in usual perturbative QCD. These involve all (light) quark
flavours democratically  along with a violation of chirality, in accord
with the general chiral anomaly relation~\cite{th}.
Based on this crucial observation, deep-inelastic scattering (DIS) at HERA
has been shown to offer a unique opportunity\cite{rs1} 
to discover such instanton-induced processes.
It is of particular importance that a theoretical prediction of both  the
corresponding rate~\cite{mrs,rs2,rs-lat} and
the characteristic event signature~\cite{rs1,qcdins,cgrs,rs3} is
possible in this hard scattering regime\footnote{For an exploratory
calculation of the instanton contribution to the gluon-structure
function, see Ref.~\cite{bb}.}. The instanton-induced cross section
turns out to be in a measurable range~\cite{rs2,qcdins}.
Crucial information on the region of validity for this important result, based
on instanton-perturbation theory, comes from a high-quality lattice
simulation~\cite{rs-lat,ukqcd}. 
Another interesting possible spin-dependent signature of instantons in DIS,
in form of a characteristic azimuthal spin asymmetry, has recently been
discussed in Ref.~\cite{os}. 

In a first part (Sect. 2), I shall review our extensive
investigations of deep-inelastic processes induced by small instantons.
This includes a ``flow-chart'' of our calculations based on
$I$-perturbation theory\cite{mrs,rs2}, an exploitation of crucial lattice
constraints\cite{rs-lat,ukqcd} and a confrontation\cite{rs3} of the recent
$I$-search results by the HERA collaborations H1 and ZEUS\cite{h1,zeus} with
our predictions. Next I shall briefly outline in Sect.~3 an ongoing project~
\cite{mp} to investigate  theoretically and phenomenologically the discovery
potential of hard instanton processes at the LHC. In Sect.~4, I then turn to
an overview of our work~\cite{fs,su1,su2,su3} on high-energy processes
involving larger-sized instantons. I shall focus mainly on the important
theoretical challenge of the phenomenon of saturation at small Bjorken-$x$
from an instanton perspective. In such a framework we
found~\cite{fs,su1,su2,su3} that the conspicuous average instanton size scale,
$\rav \sim 0.5$ fm, as known from lattice simulations~\cite{ukqcd}, plays the
r\^{o}le of the saturation scale. As a further main and intriguing result, we
were led to associate the ``Colour Glass Condensate"~\cite{cgc} with the QCD 
{\it sphaleron} state~\cite{sphal1}. For another more recent approach to
small-$x$ saturation in an instanton background with main emphasis on Wilson
loop scattering and lacking direct lattice input, see Ref.~\cite{sz}. The
conclusions of this overview may be found in Sect.~5. 

\section{Small instantons in deep-inelastic scattering}
\subsection{Instanton-perturbation theory}
\begin{figure}
  \begin{center}
    \parbox{10cm}{\includegraphics*[width=10cm]{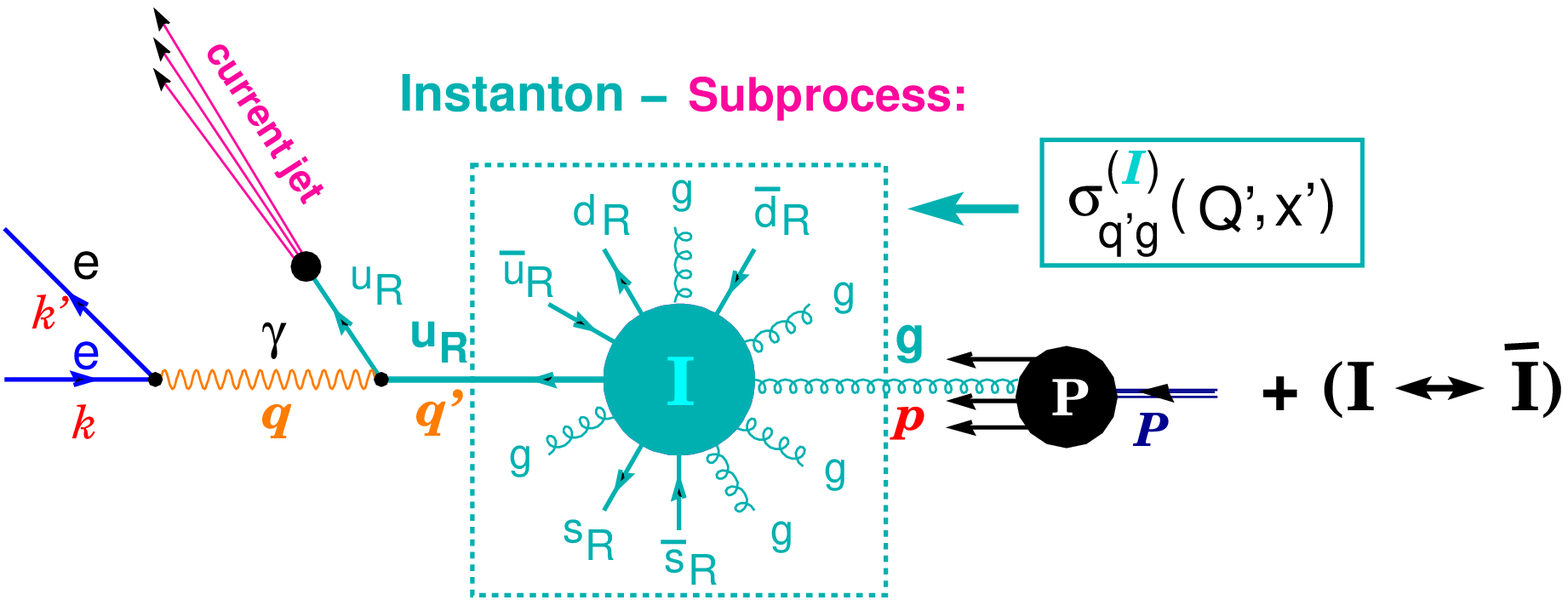}}\hfill
    \parbox{5.5cm}{\includegraphics*[width=5.5cm]{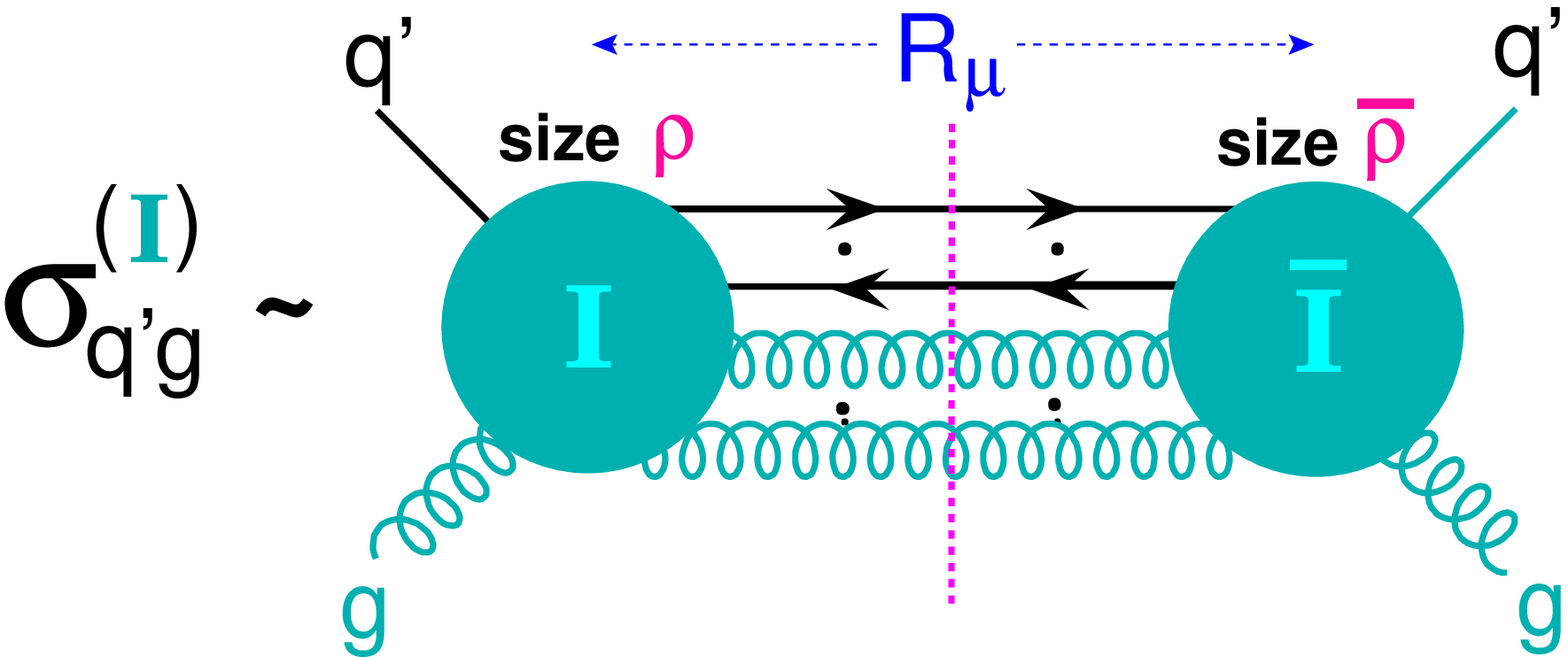}}
    \caption[dum]{\label{opt-th} (left): Leading, instanton-induced process in
deep-inelastic $e^\pm {\rm P}$ scattering for $n_f = 3$ massless flavours.
(right): 
      Structure of the total cross section
      $\sigma^{(I)}_{{  q^\prime}\,g}$ for the chirality-violating
      ``instanton-subprocess'' $q^\prime\,g \stackrel{(I)}{\Rightarrow} X$ 
      according to the optical theorem. Note the
      illustration of the collective coordinates $\rho,\overline{\rho}$
      and $R_\mu$. 
      }
  \end{center}
\end{figure}

Let us start by briefly summarizing the essence of
our theoretical calculations\cite{mrs,rs2} based on so-called
$I$-perturbation theory. As we shall see below, in an appropriate
phase-space region of deep-inelastic scattering with generic hard
scale $\mathcal{Q}$, the contributing $I$'s and $\overline{I}$'s have {\it
small  size} $\rho\lwig\mathcal{O}(\frac{1}{\alpha_s(\mathcal{Q})\mathcal{Q}})$
and  may be self-consistently considered as a {\it dilute} gas, with the small
QCD  coupling $\alpha_s(\mathcal{Q})$ being the expansion parameter like in
usual  perturbative QCD (pQCD). Unlike the familiar expansion about 
the trivial vacuum $A^{(0)}_\mu=0$ in pQCD, in $I$-perturbation theory   
the path integral for the generating functional of the   
Green's functions in Euclidean position space is then expanded about the
known, classical one-instanton solution, $A_\mu=A^{(I)}_\mu(x)+\ldots$.
After Fourier transformation to momentum space, LSZ amputation and
careful analytic continuation to Minkowski space (where the actual
on-shell limits are taken), one obtains a corresponding set of modified
Feynman rules for calculating $I$-induced scattering amplitudes.  
As a further prerequisite, the masses $m_q$ of the active quark
flavours must be light on the scale of the inverse effective $I$-size
$1/\rho_{\rm eff}$, i.\,e. $ {\rm m}_q\cdot\rho_{\rm eff}\ll 1$.
The leading, $I$-induced, chirality-violating process in the deep-inelastic
regime of $e^\pm {\rm P}$ scattering is displayed in Fig.~\ref{opt-th}\,(left)
for $n_f=3$ $~$ massless flavors. In the background of an $I\ (\overline{I}\,)$
(of  topological charge $Q=+1\ (-1)$), all $n_f$ massless quarks and
anti-quarks are  right (left)-handed such that the $I$-induced subprocess
emphasized in the dotted box of Fig.~\ref{opt-th}\,(left) involves a violation
of chirality 
$Q_5=\#\,(q_{\rm
  R}+\overline{q}_{\rm R})-\#\,(q_{\rm L}+\overline{q}_{\rm L})$ by an amount,  
\begin{equation}
  \Delta Q_5 =2\,n_f\,Q, 
\end{equation}
in accord with the general chiral anomaly relation\cite{th}.  
Within $I$-perturbation theory, one first of all derives
the following factorized expression in the Bjorken limit of
the $I$-subprocess variables $Q^{\prime\,2}$ and $\xpr$ (c.\,f.
Fig.~\ref{opt-th}\,(left)): 
\begin{equation}
  \frac{{\rm d}\sigma_{\rm  HERA}^{({ I})}}{{\rm d}{\xpr} {\rm
      d}{ Q^{\prime\,2}}}\simeq \frac{ {\rm d}{\cal
      L}_{{ q^\prime} g}^{({ I})}}{{\rm d}{\xpr} {\rm
      d}{ Q^{\prime\,2}}}\cdot { 
    \sigma_{{ q^\prime} g}^{({ I})}({ Q^\prime},{\xpr})}
  \hspace{3ex}{\rm for\ }\left\{ 
    \begin{array}{l} 
      Q^{\prime\,2}=-q^{\prime\,2}>0 { \rm \ large},\\  
      0\le \xpr=\frac{Q^{\prime\,2}}{2 p\cdot q^\prime}\le 1 {\rm \ fixed\ }.
    \end{array} 
  \right .
  \label{Bjlim}
\end{equation}
In Eq.~(\ref{Bjlim}), the differential luminosity, ${\rm d}{\cal
  L}_{q^\prime\,g}^{( I)}$ counts the number of $q^\prime\,  g$
collisions per $eP$ collisions. It is  given in terms of integrals over
the gluon density, the virtual photon flux, and the (known) flux of
the virtual quark $q^\prime$ in the instanton background\cite{rs2}. 

The essential instanton dynamics resides, however,  in the total
cross-section of the $I$-subprocess $q^\prime\ g
\stackrel{I}{\Rightarrow} X$ (dotted box of
Fig.~\ref{opt-th}\,(left) and Fig.~\ref{opt-th}\,(right)). Being an observable,
$\sigma_{q^\prime g}^{(I)}( Q^\prime, \xpr)$ involves 
integrations over all $I$ and $\overline{I}\,$-``collective coordinates'',
i.\,e. the $I\ (\overline{I}\,)$ sizes  $\rho\ (\overline{\rho}\,)$, 
the $\iai$ distance four-vector  $R_\mu$ and the relative $\iai$ color
orientation matrix $U$.  

\begin{equation}
  \sigma^{(I)}_{{  q^\prime}\,g}=
  \int d^4 { R}\ 
  {\rm e}^{\ii\, (p+{ 
      q^\prime})\cdot { R}} 
  \int\limits_0^\infty d{ \rho}
  \int\limits_0^\infty d{ \overline{\rho}}\
  {{\rm e}^{-({\rho +\overline{\rho}}){ Q^\prime}}\,}
  \ {
    D({ \rho})\, D(\overline{\rho}\,)}
  \int d{ U}\,{\rm e}^{{-\frac{4\pi}{\alpha_s}}\,
    { \Omega\left({
          U},{\frac{ R^2}{\rho\overline{\rho}}},{ 
          \frac{\overline{\rho}}{\rho}} \right)}}\,\{\ldots\}\,
  \label{cs}
\end{equation}
Both instanton and anti-instanton degrees of freedom enter here, 
since the I-induced cross-section results from taking the modulus squared of an
amplitude in the single $I$-background. Alternatively and more conveniently
(c.\,f. Fig.~\ref{opt-th}\,(right)), one may invoke the optical theorem to
obtain the cross-section (\ref{cs}) in Minkowski space as a discontinuity of
the $q^\prime\,g$ forward elastic scattering amplitude in the
$\iai$-background~\cite{rs2}. The $\{\ldots\}$ in Eq.~(\ref{cs}) abreviates
smooth contributions associated with the external partons etc.
Let us concentrate on two crucial and strongly varying quantities of the
$I$-calculus appearing in Eq.~(\ref{cs}): $D(\rho)$, the (reduced) $I$-size
distribution\cite{th,bernard}, and
$\Omega\left(U,\frac{R^2}{\rho\overline{\rho}}, 
  \frac{\overline{\rho}}{\rho} \right)$, the $\iai$
interaction, associated with a resummation of
final-state gluons. Both objects are {\it known} within
$I$-perturbation theory, formally for  $\alpha_s(\mu_r)\ln(\mu_r\,\rho)\ll 1$
and $\frac{R^2}{\rho\overline{\rho}} \gg 1$ (diluteness), respectively,
with $\mu_r$ being the renormalization scale. In the $\iai$-valley
approach~\cite{yung}, the functional form of $\Omega_{\rm valley}^{I\bar{I}}$ is
analytically known~\cite{kr,verbaarschot} (formally) for {\em all}
values of $R^2/(\rho\bar{\rho})$. The {\it actual} region of validity of the
valley approach is an important issue to be addressed again later.
 
Most importantly, the resulting power-law behaviour for the $I$-size
distribution,
\begin{equation}
  D(\rho)\propto \rho^{\beta_0-5 +\mathcal{O}(\alpha_s)}, 
\end{equation} 
involving the leading QCD $\beta$-function coefficient,
$\beta_0=\frac{11}{3}\,N_c-\frac{2}{3}\,n_f,\ (N_c=3)$, 
generically spoils the calculability of $I$-observables due to the
bad IR-divergence of the integrations over the $I\
(\overline{I}\,)$-sizes   for large $\rho\ (\overline{\rho}\,)\,$. 
Deep-inelastic scattering represents, however, a crucial exception: 
The {\it exponential} ``form factor''
$\exp({- Q^\prime}(\rho+\overline{\rho}\,))$ that was shown\cite{mrs}
to arise in Eq.~(\ref{cs}), insures convergence and {\it small} instantons
for large enough $Q^\prime$, despite the strong power-law growth of
$D(\rho)$. 
This is the key feature, warranting the calculability of
$I$-predictions for DIS.

It turns out that for (large) $Q^\prime \ne 0$, all collective
coordinate integrations in $\sigma_{q^\prime g}^{(I)}$ of Eq.~(\ref{cs}) may
be performed in terms of a {\it unique saddle point}: 
\begin{eqnarray}
  U^\ast&\Leftrightarrow& \mbox{\rm \ most attractive relative $\iai$
    orientation in color space},\nonumber\\ 
  \rho^\ast &=& \overline{\rho}^\ast\sim
  \frac{4\pi}{\alpha_s(\frac{1}{\rho^\ast})}\,\frac{1}{ Q^\prime};    
  \hspace{2ex}
  \frac{
    R^{\ast\,2}}{\rho^{\ast\,2}} \stackrel{ \ Q^\prime {\rm\ large}\
  }{{\sim}} 4\frac{\xpr}{ 1-\xpr}   
  \label{saddle}
\end{eqnarray}
This result underligns the self-consistency of the approach, since for
large $Q^\prime$ and  small $(1-\xpr)$ the  
saddle point (\ref{saddle}), indeed, corresponds to widely separated, small
$I$'s and $\overline{I}$'s.  

\subsection{Crucial impact of lattice results}

The $I$-size distribution $D(\rho)$ and the $\iai$ interaction $\Omega\left(
  U,\frac{ R^2}{\rho\overline{\rho}}, \frac{\overline{\rho}}{\rho} \right)$
form a crucial link between deep-inelastic scattering and 
lattice observables in the QCD vacuum\cite{rs-lat}. 

Lattice simulations, on the other hand, provide independent,
non-perturbative information on the 
{\it actual} range of validity of the form predicted from $I$-perturbation
theory for these important functions
of $\rho$ and $R/\rho$, respectively. The one-to-one saddle-point
correspondence (\ref{saddle}) of  the (effective) collective
$I$-coordinates ($\rho^\ast,R^\ast/\rho^\ast$) to $(Q^\prime, \xpr)$
may then be exploited to arrive   
at a ``fiducial'' $(Q^\prime, \xpr)$ region for our predictions in
DIS. Let us briefly summarize the results of this strategy\cite{rs-lat}.

We have used the high-quality lattice
data\cite{ukqcd,rs-lat} for quenched QCD ($n_f=0$) by the UKQCD
collaboration together with the careful, non-perturbative lattice
determination of the respective QCD $\Lambda$-parameter, 
$\Lambda^{(n_f=0)}_{\overline{\rm MS}}= (238{\pm  19}) {\rm \ MeV}$,
by the ALPHA collaboration\cite{alpha}.  
\begin{figure}
  \begin{center}   
   \parbox{6cm}{\vspace{0.5ex}\includegraphics*[width=6cm]{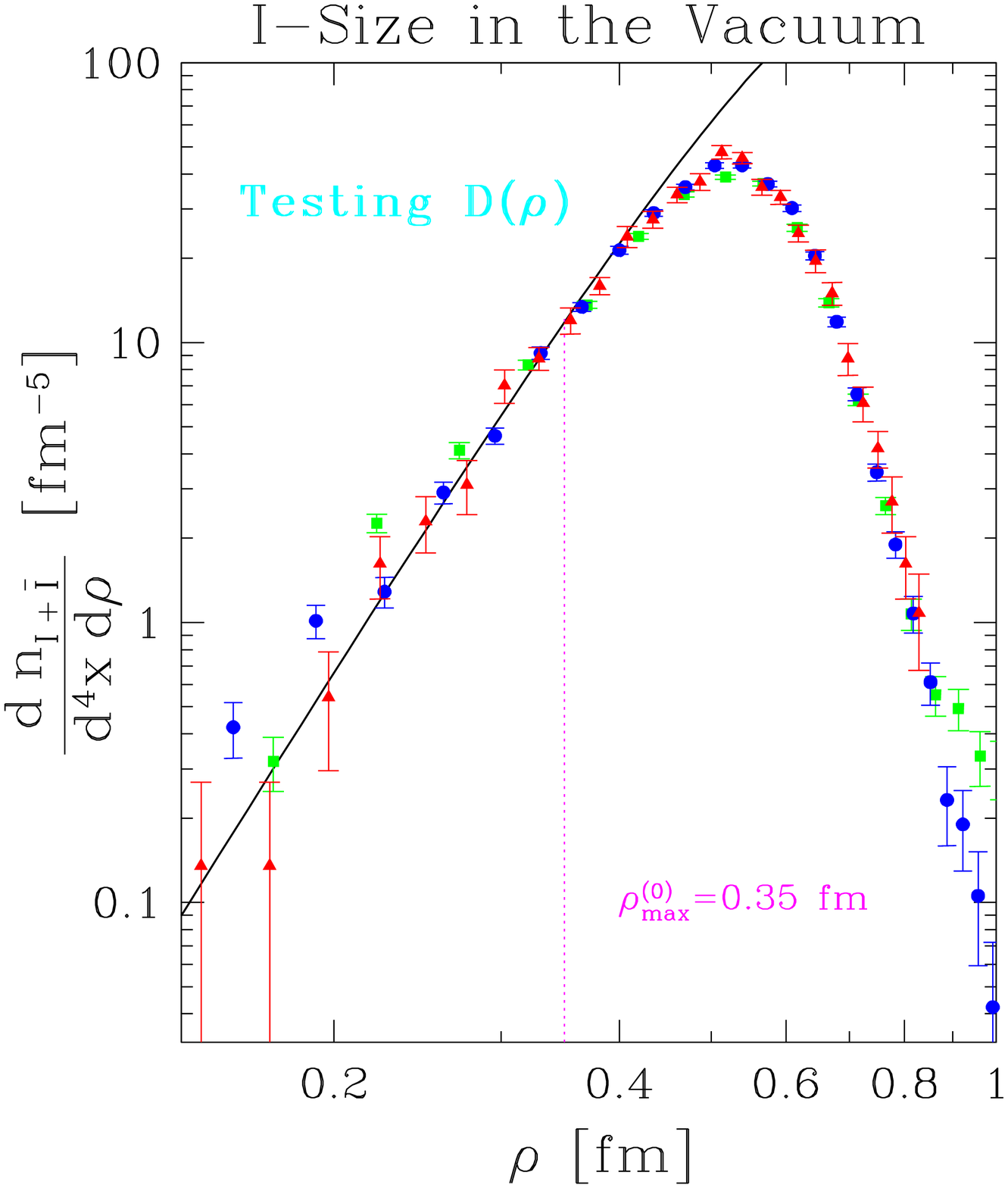}\vspace{
0.5cm}
    }\hspace{0.8cm}\parbox{6cm}{\includegraphics*[width=6cm]{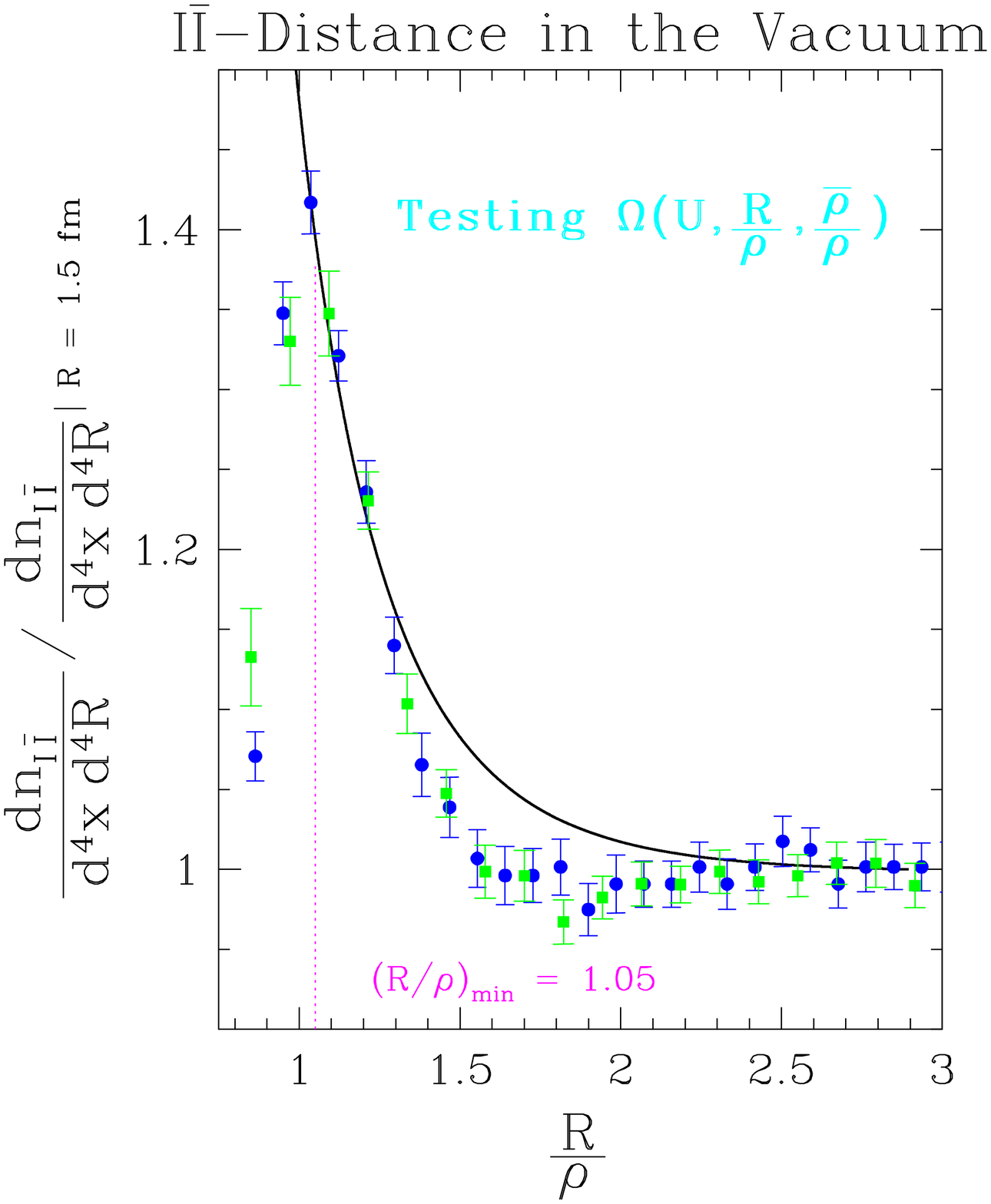}}
    \end{center}
  \vspace{-4ex}  
  \caption[dum]{\label{lattice} Illustration of the agreement
    of recent high-quality lattice data~\cite{ukqcd,rs-lat} for the
    instanton-size distribution (left) and the normalized $I\overline{I}$-distance
    distribution (right) with the predictions from  
    instanton-perturbation theory~\cite{rs-lat} for $\rho\lwig
    0.35$ fm and $R/\rho\gwig 1.05$, respectively. 
    $\alpha_{\rm \overline{MS}}^{\rm 3-loop}$ with $\Lambda_{{\rm
        \overline{MS}}}^{(n_f=0)}$ from the ALPHA collaboration~\cite{alpha} was
    used.}
\end{figure}
The results of an essentially
parameter-free comparison of the continuum limit\cite{rs-lat} for the
simulated $(I+\overline{I})$-size and the 
$\iai$-distance distributions with $I$-perturbation theory versus
$\rho$ and $R/\rho$, respectively, is displayed in Fig.~\ref{lattice}.
The UKQCD data for the $\iai$-distance distribution provide the first
direct test of the $\iai$ interaction $\Omega\left( U,\frac{
    R^2}{\rho\overline{\rho}}, \frac{\overline{\rho}}{\rho} \right)$
from the $\iai$-valley approach via\cite{rs-lat}  
\begin{eqnarray} 
  \frac{{\rm d}\,n_{I\overline{I}}}{{\rm d}^4x\, {\rm d}^4 R}_{|\rm UKQCD} 
  \stackrel{?}{\simeq}
  \int\limits_0^\infty d\,\rho\,
  \int\limits_0^\infty d\,\overline{\rho}\,
  D(\rho)\, D(\overline{\rho})
  \int d\,U\,{\rm e}^{-\frac{4\pi}{\alpha_s}\,
    \Omega\left(U,\frac{R^2}{\rho\overline{\rho}},\frac{\overline{\rho}}{\rho}
    \right)},   
\end{eqnarray}
and the lattice measurements of $D(\rho)$.

From Fig.~\ref{lattice}, $I$-perturbation theory appears to be
quantitatively valid for 
\begin{equation}
  \left.\begin{array}{lcl}{\rho\cdot\Lambda^{(n_f=0)}_{\overline{\rm
            MS}}}&{ \lwig}&{ 0.42}\\[1ex]{ R}/{ \rho}&{\gwig}&{ 1.05}\\ 
    \end{array}
  \right\}\stackrel{\rm\bf saddle\ point}{\Rightarrow}
  \left\{\begin{array}{lcl}{ Q^\prime}/
      {\Lambda^{(n_f)}_{\overline{\rm MS}}}&{ \gwig}&{ 
        30.8},\\[2ex] 
      { x^\prime}&{ \gwig}&{  0.35},\\\end{array} \right .
  \label{fiducial}
\end{equation}
Beyond providing a quantitative estimate for the ``fiducial'' momentum
space region in DIS, the good, parameter-free  agreement of the
lattice data with $I$-perturbation theory is very interesting in its own
right.
Uncertainties associated with the inequalities (\ref{fiducial}) are studied in
detail in Ref.~\cite{rs3}.

\subsection{Characteristic final-state signature}
\begin{figure}[ht]
  \begin{center}
    \parbox{4.5cm}{\includegraphics*[width=4.2cm]{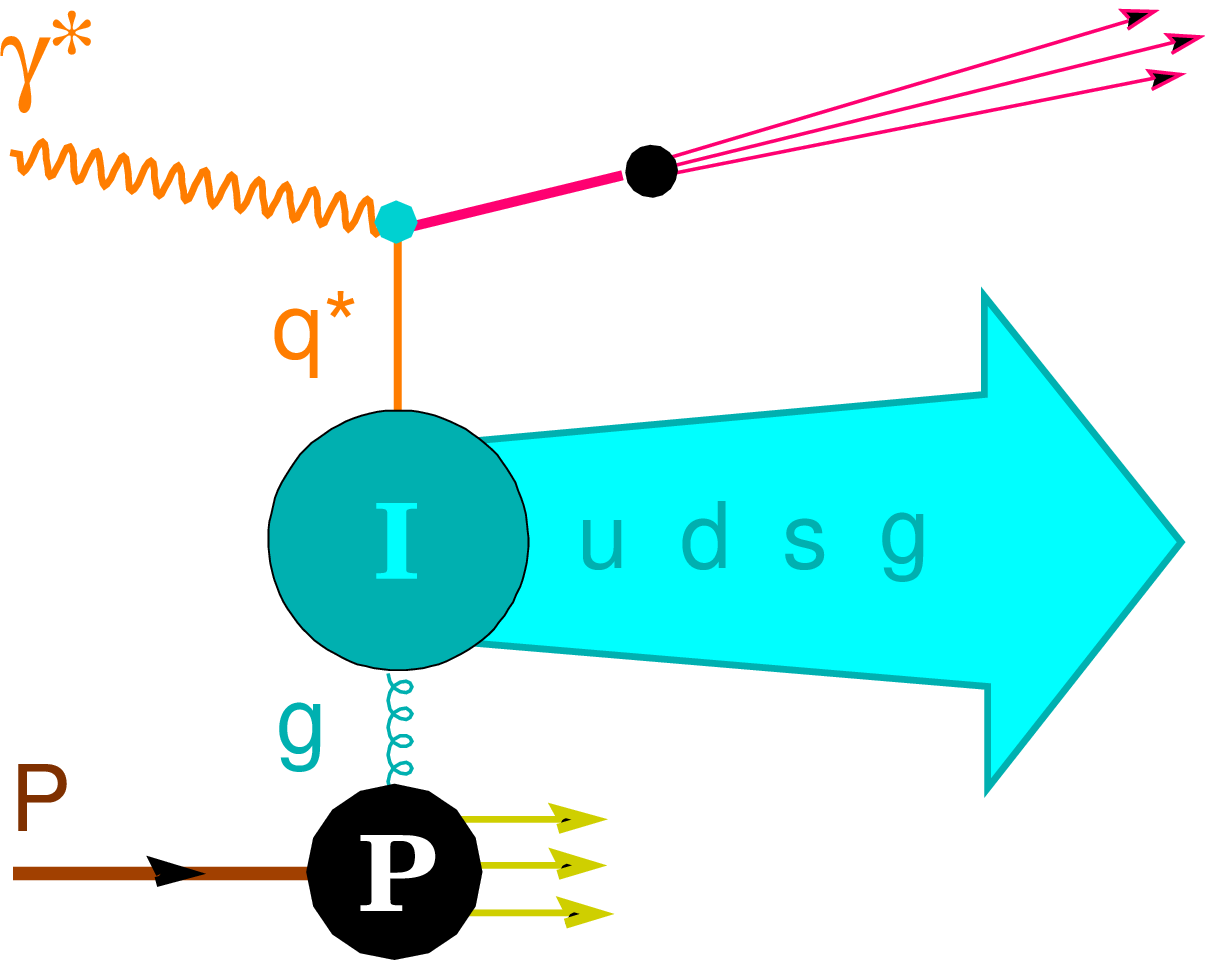}}
    \parbox{6.2cm}{\vspace{-4.5ex}\begin{minipage}{5.6cm}{
          \small current {\bf jet}\\[0.8cm]
          {\bf ``band''}-region:
          ``Fireball'' decaying {\it
            isotropically} (in $I$-rest system) into    
          $n_f\,(q +\overline{q}\,)+\,
          {{\cal O}(\frac{ 1}{\alpha_s})}\,
          g=\mathcal{O}(10)$ partons  
        }\end{minipage}}
    \parbox{5cm}{\includegraphics*[width=5cm]{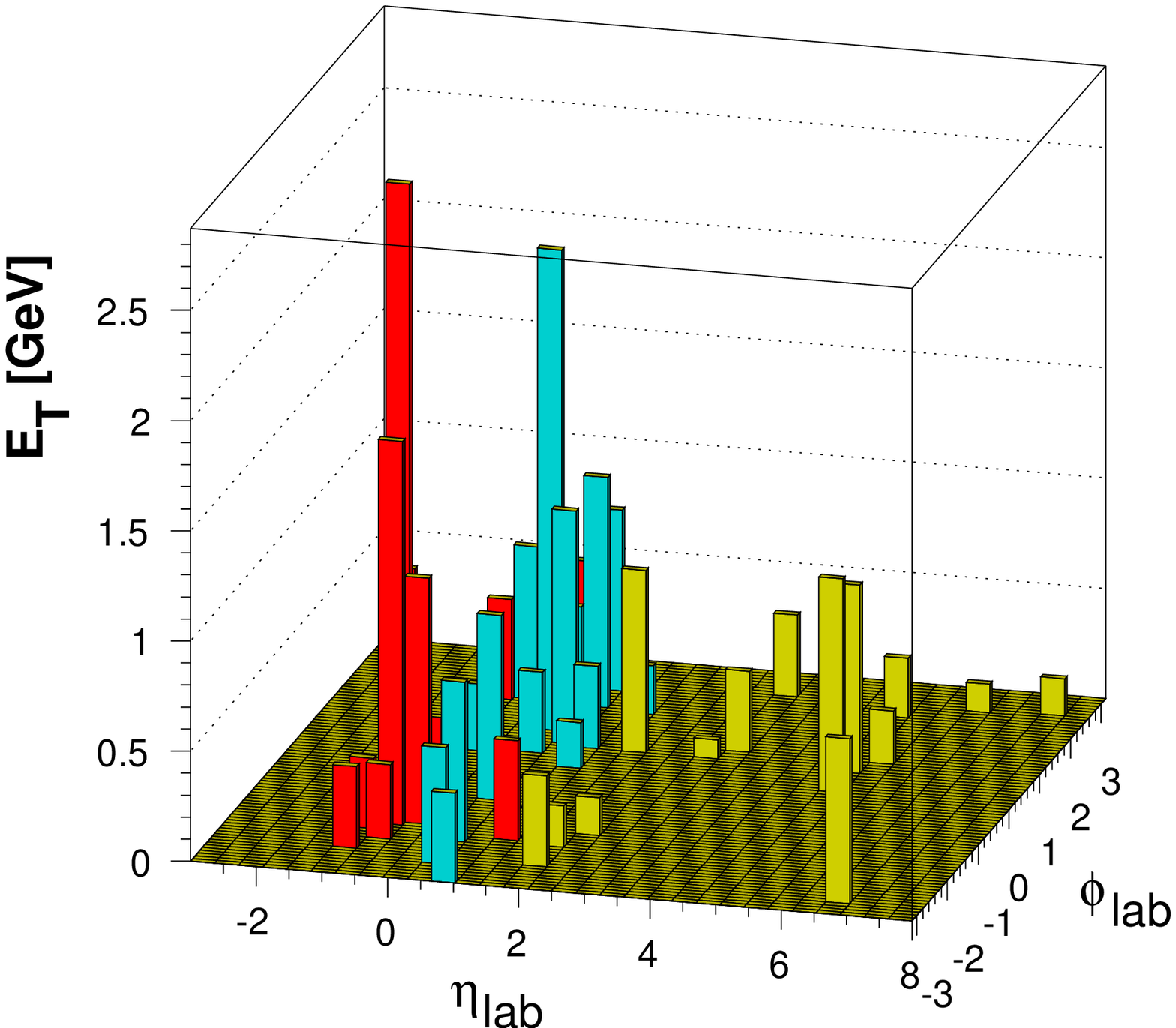}}
  \end{center}
  \caption[dum]{\label{event}Characteristic signature of $I$-induced
    events: {\it One} (current) {\it jet} along with a densely filled {\it
      band} of hadrons in the ($\eta,\phi$) plane. Each event has large
    hadron multiplicity, large total $E_t$, u-d-s flavor democracy
    with 1 $s\overline{s}$-pair/event leading to $K's,\Lambda's\ldots$. An
    event from our QCDINS\cite{qcdins} generator (right) illustrates
    these features.} 
\end{figure}

The qualitative origin of the characteristic final-state signature of
$I$-induced events is intuitively explained and illustrated in
Fig.~\ref{event}.
An indispensable tool for a quantitative investigation of the characteristic
final-state signature and notably for actual experimental searches of
$I$-induced events at HERA is our  Monte-Carlo generator package
QCDINS\cite{qcdins}. 

\subsection{Status of searches at HERA}
The results of dedicated searches for instanton-induced events by the H1 and
ZEUS collaborations~\cite{h1,zeus}, based on our theoretical work, have been
finalized meanwhile. The H1 analysis was based on $\int \mathcal{L} dt \approx
21$ pb$^{-1}$, while ZEUS used $\int \mathcal{L} dt \approx
38$ pb$^{-1}$, with somewhat differing kinematical cuts.
Since the upgraded HERA II machine is now performing
very well, forthcoming  searches based on a several times higher luminosity
might turn out most interesting. Let me briefly summarize the present
status from a theorist's perspective.

While H1 indeed observed a statistically significant excess of events with
instanton-like topology and in good agreement with the theoretical predictions,
{\it physical} significance could not be claimed, due to remaining uncertainties
in the standard DIS (sDIS) background simulation. The ZEUS collaboration
obtained a conservative, background-independent upper limit on the
instanton-induced HERA cross section of $26\ {\rm pb}@95\%{\rm\ CL}$, to be
compared to our theoretical prediction of 8.9 pb for the given cuts. In both
experiments it was demonstrated that a decisive experimental test of the
theoretical predictions based on I-perturbation theory is well within reach in
the near-future. In view of the present situation and the interesting prospects
for HERA II, let me proceed with a number of comments.

A first important task consists in reconstructing the instanton-subprocess
variables ($Q^{\prime\,2}, x^\prime$) from Eq.~(\ref{Bjlim}) and in implementing
the theoretically required fiducial cuts (cf. Eq.~(\ref{fiducial})). The actual
status is displayed in Table~\ref{cuts} for comparison.
\begin{table}
\begin{center}
\begin{tabular}{|lcl||c|c|}\hline 
Fiducial        &   Cuts      &                 & {\bf H1}&{\bf
ZEUS}\\\hline\hline
$Q^2$           &  $\gwig$ & $113$ GeV$^2$ ? & {\bf no} & {\bf yes}\\\hline
$Q^{\prime\,2}$ &  $\gwig$ & $113$ GeV$^2$ ? & {\bf yes}& {\bf yes}\\\hline
$x^\prime$      &  $\gwig$ & $0.35$        ? & {\bf no} & {\bf no} \\\hline
\end{tabular}
\caption[dum]{\label{cuts} Comparison of implemented fiducial cuts that are
required in principle to warrant the validity of I-perturbation theory.}
\end{center}
\end{table} 
The implications of the lacking $x^\prime$-cut both in the H1 and ZEUS data
are presumably not too serious, since QCDINS --{\it with} its default
$x^\prime$-cut-- models to some extent the sharp suppression of
$I$-effects, apparent in the lattice data (cf. Fig.~\ref{lattice}\,(right))
for
$ R/\rho\lwig 1.0-1.05$, i.e. $ x^\prime \lwig 0.3-0.35$.     
Yet, this lacking, experimental cut introduces a substantial
uncertainty  in the predicted  magnitude of the $I$-signal that hopefully
may be eliminated soon. The lacking $Q^2$-cut in the H1 data is potentially
more serious. As a brief reminder\cite{mrs,qcdins}, this cut 
assures in particular the dominance of ``planar'' handbag-type graphs in
$\sigma^{(I)}_{\rm HERA}$ and all final-state observables.  
Because of computational complications, the non-planar
contributions are {\it not} implemented in the QCDINS event generator,
corresponding  to unreliable QCDINS results for small $Q^2$.

The main remaining challenge resides in the fairly large sDIS
background uncertainties. The essential reason is that the existing Monte Carlo
generators have been typically designed and tested for kinematical regions
different from where the instanton signal is expected! Although the residual
problematics is not primarily related to lacking statistics, the near-future
availability of many more events will allow to strengthen the cuts and thus
hopefully to increase the gap between signal and background.
A common search strategy consists in producing I-enriched data samples by
cutting on several discriminating observables, each one being sensitive to
{\it different} basic instanton characteristics. An optimized set may be found
according to the highest possible
\begin{equation}
\mbox{\rm instanton separation power} =\frac{\epsilon_I}{\epsilon_{\rm
sDIS}},
\end{equation}
in terms of the sDIS and instanton efficiencies, with $\epsilon_I\gwig
5-10\%$.
\begin{figure}[t]
\begin{center}
\includegraphics*[width=5.5cm,angle=-90]{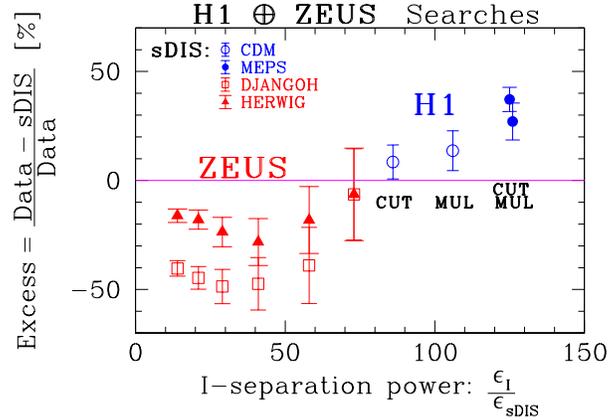}
\caption[dum]{\label{excess} A theorist's "unified
plot" of the H1 and ZEUS "excess" versus $I$-separation power. The
H1 and ZEUS data are seen to join smoothly. A first sign of a rising
excess towards higher separation powers might be suspected.}
\end{center}
\end{figure}
Substantial enhancements of the instanton sensitivity were obtained, 
by means of various multivariate discrimination methods, involving only
a single cut on a suitable discriminant variable. In case of ZEUS, cuts
on the Fisher discriminant have been used to obtain instanton-enhanced
subsamples.

Let me summarize the results obtained so far in form of a theorist's "unified
plot" of the H1 and ZEUS "excess" versus the $I$-separation power.
Any visible correlation of a rising {\it
experimental} "excess" with the (Monte-Carlo) {\it theoretical} I-separation
power in Fig.~\ref{excess} would be an intriguing first signature for a
signal. The behaviour seen from the end of the ZEUS range into the H1 domain,
might indeed suggest some increase of the excess towards rising 
I-sparation power. The comparatively low $I$-separation power of the ZEUS data
(and thus perhaps also their {\it negative} excess?) is mainly due to the
implementation of the fiducial cut in $Q^2$ that is lacking in case of H1. 

\section{Study of the discovery potential at the LHC}
Given our extended experience with instanton physics both theoretically and
experimentally at DESY, it is natural to ask about the discovery potential for
instanton-induced processes at the forthcoming LHC. Indeed, a respective project
has been set up around a theoretical PhD Thesis~\cite{mp},
but is still in a relatively early stage. 
\subsection{Outline of the project}
We attempt to do a broad study, focussing both on theoretical and
phenomenological issues. Let me just enumerate some interesting aspects
that differ essentially from the familiar situation for spacelike hard
scattering in DIS at HERA.

{\it Theoretically}: The first and foremost task is to identify and calculate
the leading $I$-subprocess at the LHC within $I$-perturbation theory. Unlike
HERA (Fig.~\ref{opt-th}\,(left)), one starts from a $g\,g$-initial
state at the LHC. Hence, the rate will be enhanced by a factor  $\propto
\frac{1}{\alpha_{\rm e.m.}\,\alpha_s}$ compared to $\gamma^\ast\,g$ scattering
at HERA. Then, the next crucial question is how to enforce some parton
virtuality in the respective instanton-induced $g\,g$-subprocess, such as
to retain the applicability of  $I$-perturbation theory.
 
An interesting possibility we are exploring is to enter the required virtuality
through the {\it final state} in case of the LHC! One may consider the
fragmentation of one or even two outgoing quarks from the $g\,g$-initiated
$I$-instanton subprocess into a {\it large} $E_\perp$ photon or $W$-boson and
other particles. The requirement of large $E_\perp$ then enforces a {\it
timelike} virtuality onto the outgoing parent quark. 

{\it Experimentally}: Crucial criteria will be a
good signature paired with the lowest possible background, as well as a good
trigger. At the experimental front we forsee the collaboration of T.
Carli/CERN, who will be able to merge his actual knowledge of the LHC with
many years of experience from searches for instantons at HERA. After the
theoretical calculations are under control, the next task is to adapt our
QCDINS event generator to the LHC, to work out characteristic event
signatures, optimal observables, fiducial cuts etc. 
\section{Instanton-driven saturation at small $x$}

One of the most important observations from HERA is the strong rise
of the gluon distribution at small Bjorken-$x$~\cite{HERA}. On the one
hand, this rise is predicted by the DGLAP evolution
equations~\cite{DGLAP} at high $Q^2$ and thus supports
QCD~\cite{riseDGLAP}. On the other hand, an undamped rise will eventually
violate
unitarity. The reason for the latter problem is known to be
buried in the linear nature of the DGLAP- and the
BFKL-equations~\cite{BFKL}: For decreasing Bjorken-$x$, the
number of partons in the proton rises, while their effective size
$\sim 1/Q$ increases with decreasing $Q^2$. At some characteristic
scale $Q^2 \approx Q_s^2(x)$, the gluons in the proton start to
overlap and so the linear approximation is no longer applicable;
non-linear corrections to the linear evolution
equations~\cite{gribovnu} arise and become 
significant, potentially taming the growth of the gluon
distribution towards a  ``saturating'' behaviour.

From a theoretical perspective, $eP$-scattering at small Bjorken-$x$ and
decreasing $Q^2$ uncovers a novel regime of QCD, where the coupling
$\alpha_s$ is (still) small, but the parton densities are so large
that conventional perturbation theory ceases to be applicable, eventually.
Much interest has recently been generated through association of the
saturation phenomenon with a multiparticle quantum state of high
occupation numbers, the ``Colour Glass Condensate'' that
correspondingly, can be viewed~\cite{cgc} as a strong {\em classical}
colour field $\propto 1/\sqrt{\alpha_s}$.

\subsection{Why instantons?}

Being extended non-perturbative fluctuations of the gluon
field, instantons  come to mind naturally in the context of saturation, since 
\begin{itemize}
\item classical {\em non-perturbative} colour fields are physically
  appropriate in this regime; $I$-interactions always involve many
  non-perturbative gluons with  multiplicity $\langle n_g\rangle\propto
  \frac{1}{\alpha_s}$!
\item the functional form of the instanton gauge field is explicitely
  known and its strength is $A_\mu^{(I)}\propto \frac{1}{\sqrt{\alpha_s}}$ as
  needed;
\item an identification of the ``Colour Glass Condensate'' with the
  QCD-sphaleron state appears very suggestive~\cite{su2,su3}
  (cf. below and Sec~4.4).
\item At high energies ($x\rightarrow 0$), larger $I$-sizes ($\rho\gwig 0.35$
 fm) are probed! Unlike DIS, now the sharply defined average
$I$-size $\rav\approx 0.5$ fm (known from lattice simulations~\cite{ukqcd})
comes into play and becomes a relevant and conspicuous length scale in this
regime (cf. Fig.~\ref{lambda}\,(left)). 
\item An intriguing observation is that the $I$-size scale $\rav$ coincides
surprisingly well with the transverse resolution $\Delta x_\perp\sim 1/Q$, where
the small-$x$ rise of the structure function $F_2(x,Q^2)$ {\it abruptly} starts
to {\it increase} with falling $\Delta x_\perp$! This striking
feature\footnote{I wish to thank A. Levy for the experimental data in
Fig.~\ref{lambda}\,(right)} is illustrated in Fig.~\ref{lambda}\,(right), with
the
power $\lambda(Q)$ being defined via the ansatz
$F_2(x,Q^2)=c(Q)\,x^{-\lambda(Q)}$ at small $x$. A suggestive interpretation is
that instantons are getting resolved for $\Delta x_\perp\lwig\rav$.
\begin{figure}
\parbox{7.0cm}{\includegraphics*[width=7.0cm]{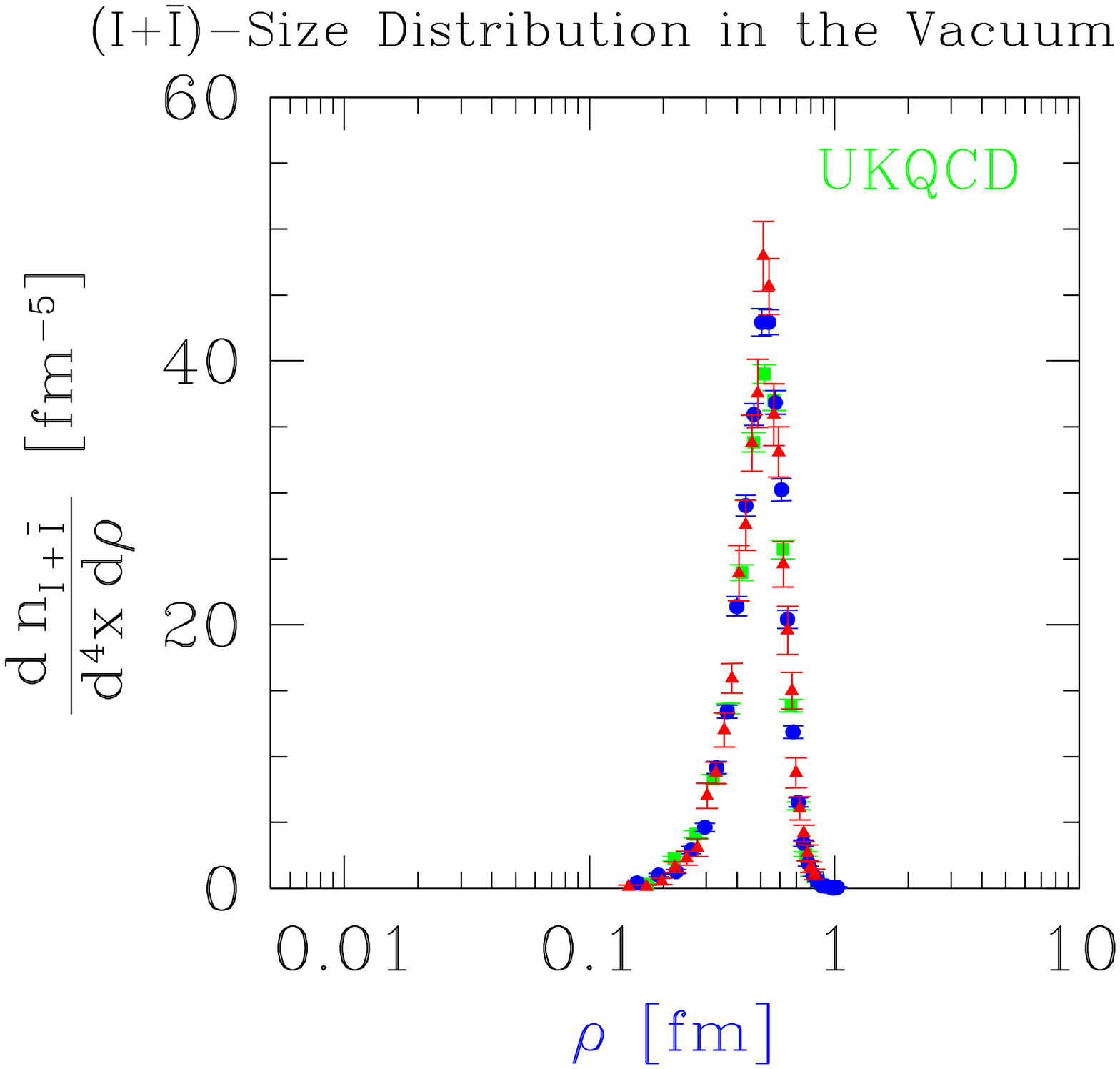}}\parbox{6.3cm}{\vspace
{
-1.5mm}\hspace{0.6cm}\includegraphics*[width=6.3cm]{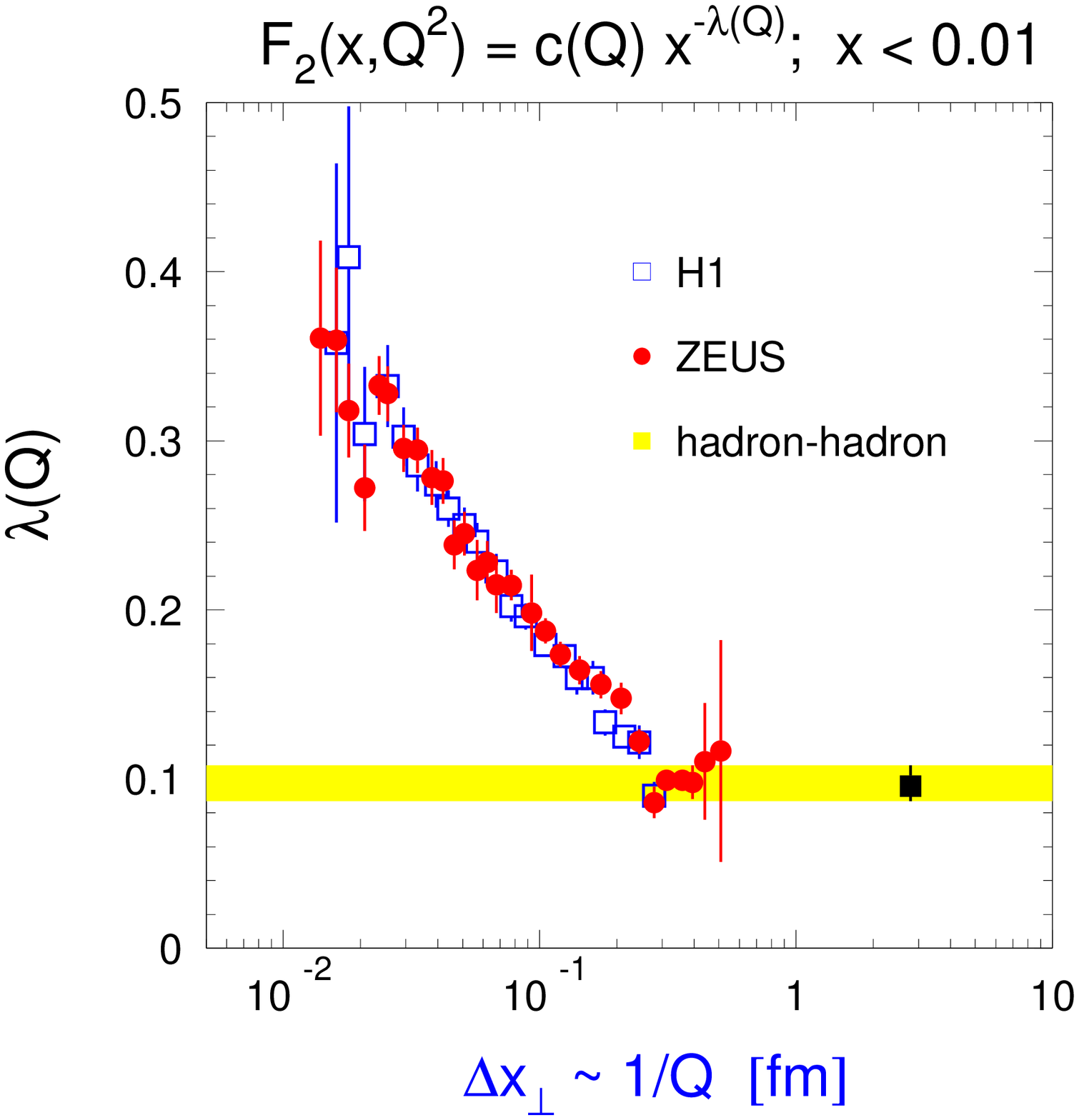}}
\caption[dum]{\label{lambda} The $I$-size scale $\rav$ from lattice data
~\cite{rs-lat,ukqcd} (left) coincides
surprisingly well with the transverse resolution \mbox{$\Delta x_\perp\sim
1/Q$}, where the small-$x$ rise of the structure function $F_2(x,Q^2)$ {\it
abruptly} starts to {\it increase} with falling $\Delta x_\perp$ (right).}
\end{figure}
\item We know already from $I$-perturbation theory that the instanton
contribution tends to strongly increase towards the softer
regime~\cite{rs1,rs2,qcdins}. The mechanism for the decreasing
instanton suppression with increasing energy is known since a long
time~\cite{sphal2,shuryak2}: Feeding increasing energy into the
scattering process makes the picture shift from one
of tunneling between adjacent vacua ($E\approx 0$) to that of the
actual creation of the sphaleron-like, coherent multi-gluon
configuration~\cite{sphal1} on top of the potential barrier of
height~\cite{rs1,diak-petrov} $E = m_{\rm
  sph}\propto\frac{1}{\alpha_s\rho_{\rm eff.}}$.
\end{itemize}
  
\subsection{From instanton-perturbation theory to saturation}

\begin{figure}
\begin{center}
\parbox{14cm}{\includegraphics*[width=14cm]{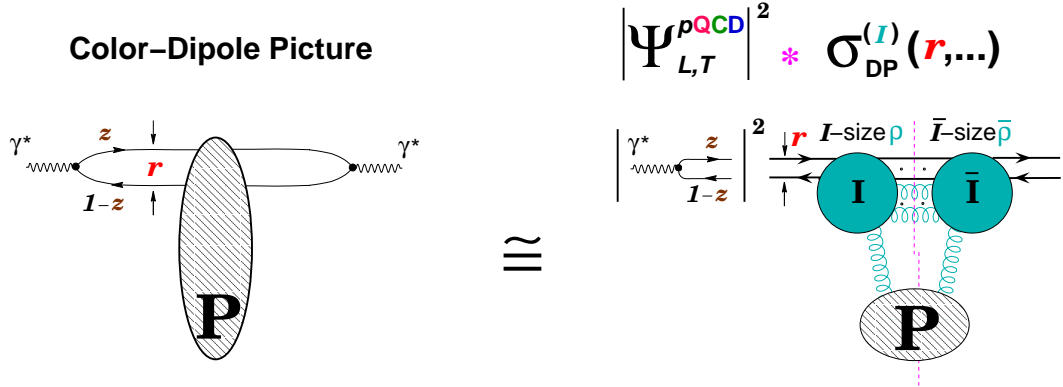}}
\end{center}
\vspace{-4ex}
\caption[dum]{\label{coldip} Illustration of the color dipole picture, its
associated variables, the factorization property and the structure of the
dipole cross section in an instanton approach. }
\end{figure}

The investigation of saturation becomes most transparent in the
familiar colour-dipole picture~\cite{dipole} (cf. Fig.~\ref{coldip}\,(left)),
notably if analyzed in the so-called dipole frame~\cite{mueller}. In this 
frame, most of the energy is still carried by the hadron, but the virtual
photon is sufficiently energetic, to dissociate before scattering into
a $q\bar{q}$-pair (a {\it colour dipole}), which then scatters off the
hadron. Since the latter is Lorentz-contracted, the dipole sees it as a colour
source of transverse extent, living (essentially) on the light cone. This colour
field is created by the constituents of the well developed hadron wave
function and -- in view of its high intensity, i.e. large occupation
numbers -- can be considered as classical. Its strength near
saturation is $\mathcal{O}(1/\sqrt{\alpha_s})$. At high energies, 
the lifetime  of the $q\overline{q}$-dipole  is much larger than the
interaction time between this $q\overline{q}$-pair and the hadron and hence,
at small $\xbj$, this gives rise to the familiar factorized
expression of the inclusive photon-proton cross sections, 
\begin{equation}
  \sigma_{L,T}(\xbj,Q^2) =\int_0^1 d z \int d^2\,r\;
  |\Psi_{L,T}(z,r)|^2\,\sigma_{\mbox{\tiny DP}}(r,\ldots).  
  \label{dipole-cross}
\end{equation}
Here, $|\Psi_{L,T}(z,r)|^2$ denotes the modulus squared of the 
(light-cone) wave function of the virtual photon, calculable in pQCD,
and $\sigma_{\mbox{\tiny DP}}(r,\ldots)$ is the
$q\overline{q}$-dipole\,-\,nucleon cross section. The variables in
Eq.~(\ref{dipole-cross}) are the transverse $(q\overline{q})$-size $\mathbf r $ 
and the photon's longitudinal momentum fraction $z$ carried by the quark. 
The dipole cross section is expected to include in general the main
non-perturbative contributions. For small $r$, one finds within
pQCD~\cite{dipole,dipole-pqcd} that $\sigma_{\mbox{\tiny DP}}$ vanishes 
with the area $\pi r^2$ of the $q\overline{q}$-dipole. Besides this phenomenon
of ``colour transparency'' for small $r=|r|$,  the dipole cross
section is expected to saturate towards a constant, once the
$q\overline{q}$-separation $r$ exceeds a certain saturation scale $r_s$ (cf.
Fig.~\ref{coldip}\,(right)). 
While there is no direct proof of the saturation phenomenon,
successful models incorporating saturation do exist~\cite{gbw}
and describe the data efficiently.

Let us outline more precisely our underlying strategy:

\begin{itemize}
\item We start from the large $Q^2$ regime and appropriate cuts
  such that $I$-perturbation theory is strictly valid. The
  corresponding, known results on $I$-induced DIS processes~\cite{mrs}
  are then transformed into the colour-dipole picture.
\item The guiding question is: Can background instantons of size
  $\sim\langle\rho\rangle$ give rise to a saturating, geometrical
  form for the dipole cross section,
  \begin{equation}
    \sidp^{(I)}(r,\ldots)\stackrel{r\gwig \langle\rho\rangle}{\sim}\pi
    \langle\rho\rangle^2.
  \end{equation}
\item With the crucial help of lattice results, the $q\bar{q}$-dipole
  size $r$ is next carefully increased towards hadronic
  dimensions. Thanks to the lattice input, IR divergencies are removed
  and the original cuts are no longer necessary.
\end{itemize}

\subsection{The simplest process: $\gamma^\ast+ g\stackrel{(I)}{\to} q_{\rm
    R}+\overline{q}_{\rm R}$}
Let us briefly consider first the simplest $I$-induced process, 
$\gamma^\ast\,g\Rightarrow q_{\rm R}\overline{q}_{\rm R}$, with one flavour
and no final-state gluons. More details may be found in Ref.~\cite{su2}.
Already  this simplest case illustrates transparently that in the presence of a
background instanton, the dipole cross section indeed saturates with a
saturation scale of the order of the average $I$-size $\rav$.  

We start by recalling the results for the total $\gamma^\ast N$ cross section
within $I$-perturbation theory from Ref.~\cite{mrs},  
\begin{eqnarray}
  \sigma_{L,T}(\xbj,Q^2)&=&
  \int\limits^1_{\xbj} \frac{d x}{x}\left(\frac{\xbj}{
      x}\right)G\left(\frac{\xbj}{x},\mu^2\right)\int d  t \frac{d
    \hat{\sigma}_{L,T}^{\gamma^* g}(x,t,Q^2)}{d t};\,\label{general}\\[2ex] 
  \frac{d\hat{\sigma}_{L}^{\gamma^* g}}{d  t}&=&\frac{\pi^7}{2}
  \frac{e_q^2}{Q^2}\frac{\alpha_{\rm em}}{\alpha_{\rm
      s}}\left[x(1-x)\sqrt{t u}\,  \frac{\mathcal{R}(\sqrt{-
        t})-\mathcal{R}(Q)}{t+Q^2}-(t\leftrightarrow  u)\right]^{\,2}, \label{mrs}
\end{eqnarray}
with a similar expression for $d\hat{\sigma}_{T}^{\gamma^\ast\,g}/d\,t$. 
Here, $G\left(\xbj,\mu^2\right)$ denotes the gluon density and $L,T$ refers to
longitudinal and transverse photons, respectively.

Note that Eqs.~(\ref{general}),~(\ref{mrs}) involve the resolution dependent
length scale 
\begin{equation}
  \mathcal{R}(\mathcal{Q})=\int_0^{\infty}
  d\rho\;D(\rho)\rho^5(\mathcal{Q}\rho)\mbox{K}_1(\mathcal{Q}\rho).
  \label{masterI}
\end{equation}
which is of key importance for continuing towards
$\mathcal{Q}\langle\rho\rangle\Rightarrow 0$!
For sufficiently large $\mathcal{Q}\langle\rho\rangle$, the crucial factor
$(\mathcal{Q}\rho)\,K_1(\mathcal{Q}\rho)\sim e^{-\mathcal{Q}\rho}$
in Eq.~(\ref{masterI}) exponentially suppresses large size instantons and
$I$-perturbation theory holds, as shown first in Ref.~\cite{mrs}.
In our continuation task towards smaller $\mathcal{Q}\langle\rho\rangle$,
crucial lattice information enters. We recall that the  $I$-size distribution
$D_{\rm lattice}(\rho)$, as {\it measured} on the
lattice~\cite{ukqcd,rs-lat,rs3}, is strongly peaked around an average $I$-size
$\langle\rho\rangle \approx 0.5$ fm, while being  
in excellent agreement with $I$-perturbation theory for $\rho \lwig 0.35$ fm
(cf. Sect.~2.2 and Fig.~\ref{lattice}\,(left)). Our strategy is thus to
generally  identify $D(\rho) = D_{\rm lattice}(\rho)$ in Eq.~(\ref{masterI}),
whence 
\begin{equation}
  \mathcal{R}(0)=\int_0^{\infty} d\rho\;D_{\rm
    lattice}(\rho)\rho^5\approx 0.3 \mbox{\ fm}
\end{equation} 
becomes finite and a $\mathcal{Q}^2$ cut is no longer necessary. 

By means of an appropriate change of variables and a subsequent $2d$-Fourier
transformation, Eqs.~(\ref{general}), (\ref{mrs}) may indeed be
cast~\cite{su2}  into a colour-dipole form (\ref{dipole-cross}), e.g. (with
$\hat{Q}=\sqrt{z\,(1-z)}\,Q$)
\begin{eqnarray}
  \lefteqn{\left(\left|\Psi_L\right|^2\sigma_{\mbox{\tiny DP}}\right)^{(I)}
    \approx\, \mid\Psi_L^{\rm pQCD}(z,r)\mid^{\,2}\,
    \frac{1}{\alpha_{\rm s}}\,\xbj\,
    G(\xbj,\mu^2)\,\frac{\pi^8}{12}}\label{resultL}\\[1ex] 
  &&\times\left\{\int_0^\infty\,d\rho
    D(\rho)\,\rho^5\,\left(\frac{-\frac{d}{dr^2}\left(2 r^2 
          \frac{\mbox{K}_1(\hat{Q}\sqrt{r^2+\rho^2/z})}{\hat{Q}\sqrt{r^2+\rho^2/z}}
        \right)}{{\rm K}_0(\hat{Q}r)}-(z\leftrightarrow 1-z)
    \right)\right\}^2.\nonumber 
\end{eqnarray} 
The strong peaking of $D_{\rm
  lattice}(\rho)$ around \mbox{$\rho\approx\rav$}, implies 
\begin{equation}
  \left(|\Psi_{L,T}|^{\,2}\sigma_{\mbox{\tiny
        DP}}\right)^{(I)}\Rightarrow\left\{\begin{array}{llcl} 
      \mathcal{O}(1) \mbox{\rm \ but exponentially small};&r\rightarrow 0,\\[2ex]
      \mid\Psi^{\rm \,pQCD}_{L,T}\mid^{\,2}\,\frac{1}{\alpha_{\rm
          s}}\,\xbj\,G(\xbj,\mu^2)\,\frac{\pi^8}{12}\,\mathcal{R}(0)^2;
      &\frac{r}{\rav}\gwig 1.\label{final} 
    \end{array}\right.
\end{equation} 
Hence, the association of the intrinsic instanton scale $\rav$ with
the saturation scale $r_s$ becomes apparent from Eqs.~(\ref{resultL}),
(\ref{final}): $\sigma_{\mbox{\tiny DP}}^{(I)}(r,\ldots)$ rises strongly as
function of $r$ around $r_s\approx\langle\rho\rangle$, and indeed {\em
  saturates} for $r/\rav>1$  towards a {\em constant
  geometrical limit}, proportional to the area
$\pi\,\mathcal{R}(0)^2\, =\,  \pi\left(\int_0^\infty\,d\rho\,D_{\rm
lattice}(\rho)\,\rho^5\right)^2$, subtended by the instanton.
Since $\mathcal{R}(0)$ would be divergent within
$I$-perturbation theory, the information about $D(\rho)$ from the 
lattice (Fig.~\ref{lambda}\,(left)) is crucial for the finiteness of the
result. 

\subsection{Identification of the color glass condensate with the
  QCD-sphaleron  state}
Next, let us consider the realistic process,
$\gamma^\ast + g\stackrel{(I)}{\to} n_f\,(q_{\rm R} +\overline{q}_{\rm R}) +
\mbox{gluons}$. On the one hand, the inclusion of final-state gluons and
$n_f>1$ causes a significant complication. On the other hand, it is due to the
effect of those gluons  that the identification of the QCD-sphaleron state
with the colour glass condensate has emerged~\cite{su2,su3}, while the
qualitative ``saturation'' features remain unchanged.  Most of the
$I$-dynamics resides in the $I$-induced $q^\ast\,g$-subprocess with an
incoming  off-mass-shell quark $q^\ast$ originating from photon dissociation.
The  important kinematical variables are the $I$-subprocess energy
$E=\sqrt{(q^\prime+p)^2}$ and the quark virtuality
$Q^{\prime\,2}=-q^{\prime\,2}$, with the gluon 4-momentum denoted by $p_\mu$.

It is most convenient to account for the final-state gluons by means of the
$I\bar{I}$-valley method~\cite{yung} (cf. also Sect.~2.1). It allows to
achieve  via the optical theorem, an elegant summation over the gluons. The
result leads  to an exponentiation of the final-state gluon effects, residing
entirely in the  $I\bar{I}$-valley interaction $-1\le\Omega_{\rm
  valley}^{I\bar{I}}(\frac{R^2}{\rho\bar{\rho}}+\frac{\rho}{\bar{\rho}}+\frac{\bar
  {\rho}}{\rho};U)\le 0\,$, introduced in Eq.~(\ref{cs}) of Sect.~2.1. Due to
the
new gluon degrees of freedom, the additional integrations over the 
$I\bar{I}$-distance $R_\mu$ appear (cf. Fig.~\ref{opt-th}\,(right)), while the
matrix $U$ characterizes the relative $I\bar{I}$ orientation in colour space.
We remember from Sect.~2.1 that the functional form of $\Omega_{\rm
valley}^{I\bar{I}}$ is analytically known~\cite{kr,verbaarschot} (formally) for
{\em all} values of $R^2/(\rho\bar{\rho})$.
\begin{figure}
  \begin{center}
    \parbox{6.1cm}{\includegraphics[width=6.1cm]{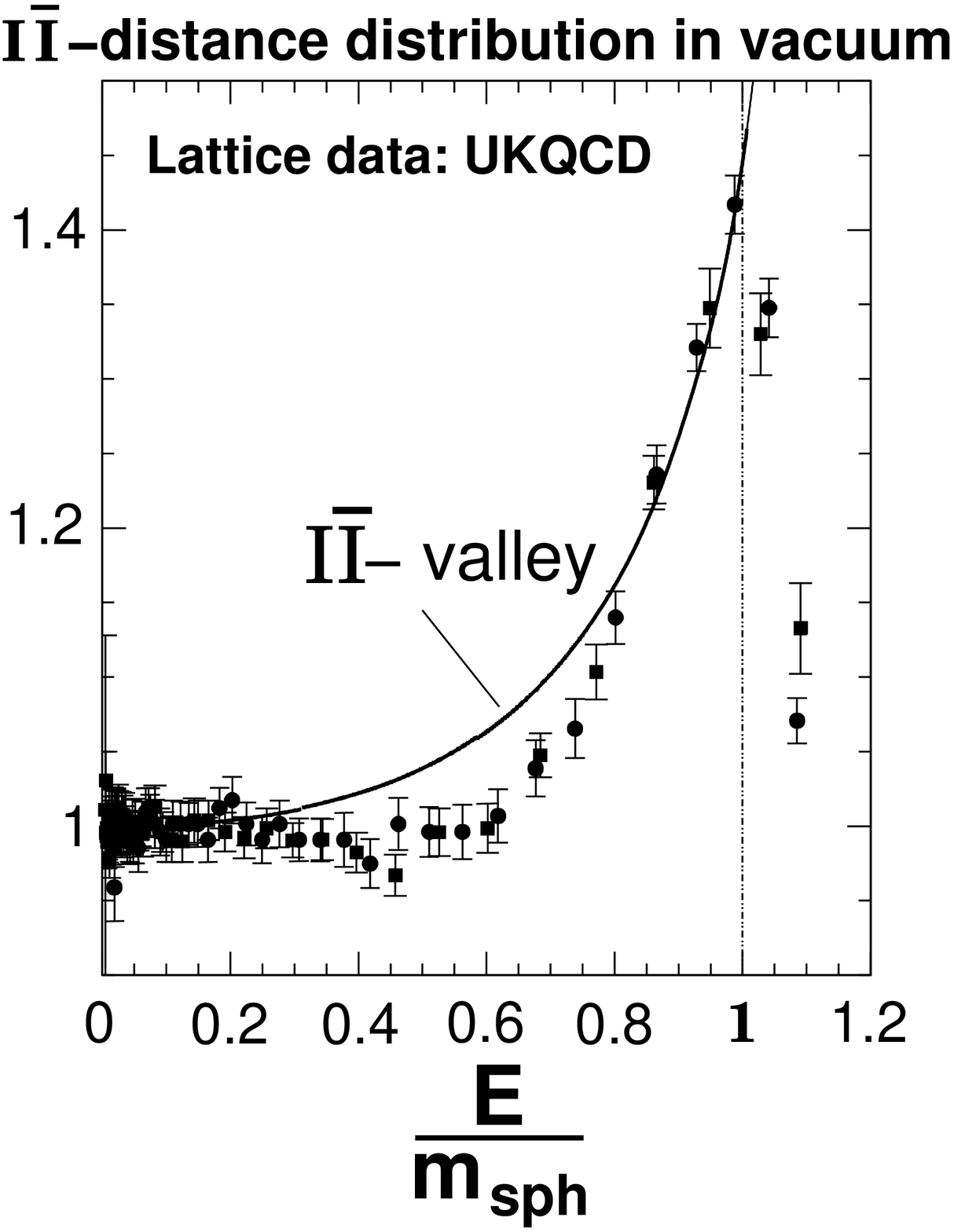}}\hspace{3ex}
    \parbox{6.9cm}{\vspace{1.5mm}\includegraphics[width=6.9cm]{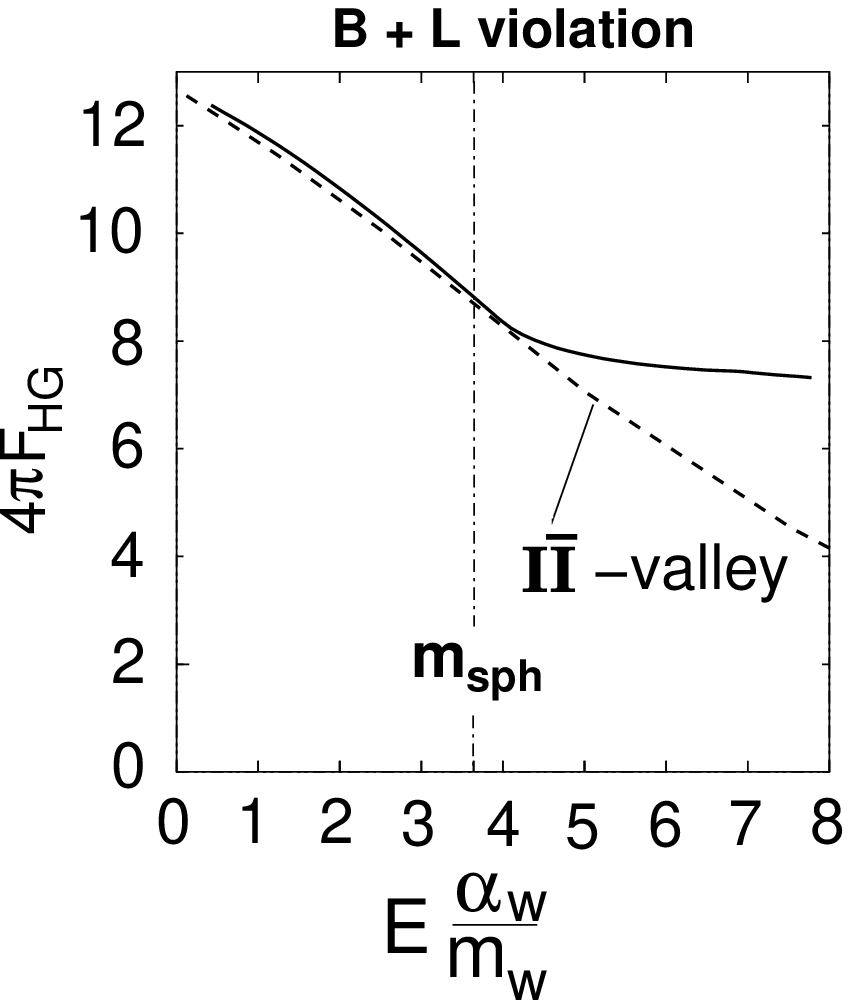}}
    \caption[dum]{
      (left) The UKQCD lattice data~\cite{ukqcd,rs-lat} of the (normalized)
      $I\bar{I}$-distance distribution together with the corresponding
      $I\bar{I}$-valley prediction~\cite{su2} from Fig.~\ref{lattice}
      (right) are re-displayed versus energy in units of the QCD
      sphaleron mass $m_{\rm sph}$.This illustrates the validity of the valley
      approach right until the sphaleron peak! 
      (right) The same trend for electroweak $B +L$ -violation is apparent from
      an independent numerical simulation of the suppression exponent for
      two-particle collisions ('Holy Grail' function) $F_{\rm HG}(E)$
      ~\cite{rubakov,ringwald}    
      \label{pic2}}  
  \end{center}
\end{figure}
Our strategy here is identical to the one for the
``simplest process'' above: Starting point is the $\gamma^\ast N$ cross
section,  this time obtained by means of the $I\bar{I}$-valley
method~\cite{rs2}. The next step is a variable and Fourier transformation into
the colour-dipole picture. The dipole cross section $\tilde{\sigma}^{(I),{\rm
    gluons}}_{\mbox{\tiny DP}}(\vl^2,\xbj,\ldots)$ before the final
    2d-Fourier transformation of the quark transverse momentum $\vl$ to
the conjugate dipole size $\vr$, arises simply as an energy integral over the
$I$-induced total $q^\ast g$ cross section in Eq.~(\ref{cs}) from
Ref.~\cite{rs2},
\begin{equation}
  \tilde{\sigma}^{(I),{\rm gluons}}_{\mbox{\tiny DP}} \approx
  \frac{\xbj}{2}\,G(\xbj,\mu^2)\,\int_0^{E_{\rm max}}
  \frac{d\,E}{E} \left[\frac{E^4}{(E^2+Q^{\,\prime 2})\,Q^{\,\prime 2}}\, 
    \sigma^{(I)}_{q^\ast \,g}\left(E,\vl^2,\ldots\right)\right],
  \label{sigdipglue}
\end{equation}
involving in turn integrations over the $I\bar{I}$-collective coordinates
$\rho,\bar{\rho},U$ and $R_\mu$.

In the softer regime of interest for saturation, we again substitute
$D(\rho) = D_{\rm lattice}(\rho)$, which enforces
$\rho\approx\bar{\rho}\approx
\langle\rho\rangle$ in the respective $\rho,\bar{\rho}$-integrals,
while the integral over the $I\bar{I}$-distance $R$ is dominated by a
{\it saddle point}, 
\begin{equation}
  \frac{R}{\rav} \approx {\rm
    function}\left(\frac{E}{m_{\rm sph}}\right); \hspace{2ex} m_{\rm
    sph}\approx \frac{3\pi}{4}\frac{1}{\alpha_{\rm s}\,\rav} =\mathcal{O}({\rm
    \,few\ GeV\,}).
  \label{sphaleron1}
\end{equation}
At this point, the mass $m_{\rm sph}$ of the
QCD-sphaleron~\cite{rs1,diak-petrov}, i.e the barrier height
separating neighbouring topologically inequivalent vacua, enters as the
scale for the energy $E$. The saddle-point dominance implies a one-to-one
relation, 
\begin{equation} 
  \frac{R}{\rav} \Leftrightarrow \frac{E}{m_{\rm sph}}; \hspace{2ex}
  \mbox{\rm with}\ R=\rav \Leftrightarrow E\approx m_{\rm sph}.
  \label{sphaleron2}
\end{equation}
Our continuation to the saturation regime now involves crucial lattice
information about $\Omega^{I\bar{I}}$. The relevant lattice observable
is the distribution of the $I\bar{I}$-distance~\cite{rs-lat,su2} $R$,
providing information on $\left\langle\exp[-\frac{4\pi}{\alpha_{\rm
s}}\Omega^{I\bar{I}}]\right\rangle_{U,\rho,\bar{\rho}}$
in euclidean space (cf. Fig.~\ref{lattice}\,(right)). Due to the crucial
saddle-point
relation Eqs.~(\ref{sphaleron1}, \ref{sphaleron2}), we may replace the
original variable $R/\rav$ by $E/m_{\rm sph}$. A comparison of the
respective $I\bar{I}$-valley predictions with the UKQCD lattice
data~\cite{ukqcd,rs-lat,su2} versus $E/m_{\rm sph}$ is displayed in
Fig.~\ref{pic2}\,(left). It reveals the important result that the
$I\bar{I}$-valley approximation is quite reliable up to $E\approx
m_{\rm sph}$. Beyond this point a marked disagreement rapidly
develops: While the lattice data show a {\it sharp peak} at $E\approx
m_{\rm sph}$, the valley prediction continues to rise indefinitely for
$E\gwig m_{\rm sph}$! It is remarkable that an extensive recent and
completely independent semiclassical numerical simulation~\cite{rubakov} shows
precisely the same trend for electroweak $B +L$-violation, as displayed in
Fig.~\ref{pic2}\,(right).

It is again at hand to identify $\Omega^{I\bar{I}}=
\Omega^{I\bar{I}}_{\rm lattice}$ for $E\gwig m_{\rm sph}$.
Then the integral over the $I$-subprocess energy spectrum  (\ref{sigdipglue})
in  the dipole cross section appears to be dominated by the sphaleron
configuration  at $E\approx m_{\rm sph}$. The feature of saturation analogously
to the  ``simplest process'' in Sect.~4.3 then implies the announced
identification of  the colour glass condensate with the QCD-sphaleron state.

\section{Conclusions}
As non-perturbative, topological fluctuations of the gluon fields, {\it
instantons} are a basic aspect of QCD. Hence their experimental discovery
through hard instanton-induced processes would be of fundamental significance.
A first purpose of this overview was to present a summary of our systematic 
theoretical and phenomenological investigations of the discovery potential in
DIS at HERA, based on a calculable rate of measurable range and a
characteristic "fireball"-like event signature. In a summary of the present
status of experimental searches by H1 and ZEUS, the typical remaining
challenges were particularly emphasized. In view of the good performance of
the upgraded HERA II machine, one may expect further possibly decisive
instanton search results in the near future.The existing H1 and ZEUS results
have demonstrated already that the required sensitivity according to our
theoretical predictions is within reach. Looking ahead, I have briefly
discussed an ongoing project concerning a broad investigation of the discovery
potential of instanton processes at the LHC. A final part of this review was
devoted to our work on small-$x$ saturation from an instanton perspective.
After summarizing the considerable motivation for the relevance of instantons
in this regime, the emerging intuitive, geometrical picture was illustrated
with the simplest example, where indeed, saturation does occur. The form of
the dipole cross section depends on the relation of two competing areas: the
area $\pi\,r^2$, subtended by the $\bar{q}q$-dipole, and the area
$\pi\,\rav^2$ associated with the average size, $\rav \approx 0.5$ fm, of the
background instanton. For $r/\rav\ll 1$, the dipole cross section is dominated
by the dipole area, corresponding to 'color transparency'. For $r/\rav\gwig 1$
it saturates towards a constant proportional to the background instanton area.
Correspondingly, the average $I$-size scale $\rav$ is associated with the
saturation scale. A further central and intriguing result concerned the
identification of the Color Glass Condensate with the QCD-sphaleron state.
Throughout, the non-perturbative information from lattice simulations was
instrumental.

\end{document}